\documentclass[journal]{IEEEtran}
\usepackage{cite}
\ifCLASSINFOpdf
\else
\fi
\usepackage{amsmath}
\usepackage{algorithmic}
\usepackage{array}
\usepackage{url}
\usepackage{mathtools}
\usepackage{amsfonts}
\usepackage{amssymb}
\usepackage{subfig}
\usepackage{graphicx}
\usepackage{wrapfig}
\usepackage{tabularx}
\usepackage{gensymb}
\usepackage[margin=1.87cm]{geometry}
\usepackage{indentfirst}
\usepackage{cite}
\usepackage{hyperref}
\usepackage{float}
\hypersetup{
    colorlinks,
    citecolor=black,
    filecolor=black,
    linkcolor=black,
    urlcolor=black
}
\usepackage{cleveref}
\usepackage{fancyhdr}
\usepackage{mathptmx}
\usepackage{subfig}
\usepackage{url}

\DeclareMathOperator*{\minimize}{{minimize}}

\begin{document}
\title{Optimization of Scalar and Bianisotropic Electromagnetic Metasurface Parameters Satisfying Far-Field Criteria}

\author{Stewart~Pearson,
        Sean~Victor~Hum,~\IEEEmembership{Senior Member,~IEEE,}% <-this % stops a space
\thanks{The authors are with the Edward S. Rogers Sr. Department of Electrical and Computer Engineering, University of Toronto, Toronto, ON M5S 3G4, Canada (e-mail: stewart.pearson@mail.utoronto.ca; sean.hum@utoronto.ca).}}% <-this % stops a space


\markboth{IEEE TRANSACTIONS ON ANTENNAS AND PROPAGATION}
{PEARSON \MakeLowercase{\textit{et al.}}: OPTIMIZATION OF SCALAR AND BIANISOTROPIC ELECTROMAGNETIC METASURFACE PARAMETERS SATISFYING FAR-FIELD CRITERIA}



\maketitle

\begin{abstract}
Electromagnetic metasurfaces offer the capability to realize almost arbitrary power conserving field transformations. These field transformations are governed by the generalized sheet transition conditions, which relate the tangential fields on each side of the surface through the surface parameters. Ideally, engineers would like to determine the surface parameters for  transformations based on their application-specific far-field criteria. However, determining the surface parameters to satisfy these criteria is challenging without direct knowledge of the tangential fields on one side of the surface, which are not unique for a given far field pattern. As a result, current design is restricted to analytical examples where the tangential fields are solvable or other \textit{ad hoc} methods. This paper presents a convex optimization-based scheme which determines surface parameters, such as surface impedance, admittance, and magneto-electric coupling, which satisfy far-field constraints such as beam magnitude, side lobe level, and null locations. The optimization is performed on a model constructed using the method of moments. This model incorporates edge effects and mutual coupling. The resulting non-convexity from this model is relaxed using the alternating direction method of multipliers. Examples of this optimization scheme performing multi-criteria pattern forming, extreme angle small surface refraction, and Chebyshev beamforming are presented.
\end{abstract}


% \fi
\IEEEpeerreviewmaketitle

\section{Introduction}

% Try to outline EMMSs a little, explain the limitations of design currently imposed (i.e. hard to go from FF criteria to ZYK, edge effect, mutual coupling), do a literature review, explain the organization of the paper.

\IEEEPARstart{E}{lectromagnetic} metasurfaces (EMMSs) are extremely powerful, two-dimensional structures capable of creating almost arbitrary wave transformations. To perform these transformations for scalar EMMSs, parameters such as surface impedance ($Z_e$) and admittance ($Y_m$) are varied spatially across the surface. More advanced transformations can be accomplished by allowing for bianisotropy, enabling tensorial $Z_e$ and $Y_m$, along with a magneto-electric coupling term $K_{em}$. These parameters relate the tangential fields directly above the surface to fields directly below the surface, which together form the desired fields on each side \cite{Kuester2003}. In practice, the surface parameters are realised by small unit cells acting like atoms in a material made of patterned conductors and dielectrics that are much smaller than the operating wavelength \cite{EpsteinOptics2016}. Because these unit cells are much smaller than the wavelength of the incident waves, they behave as though the material is homogeneous, enabling these wave transformations. Unfortunately, the current design methodology for these surfaces parameters is typically \emph{ad hoc} and relies heavily on heuristics. This major hurdle restricts EMMS design to realizing patterns that can be derived analytically. They also typically rely on model simplifications introducing inefficiencies. This is significantly holding back the potential of this exciting technology. While the design of unit cells to realize the surface parameters is a subject of intense research as well \cite{Hsu2017,Wong2014,CapekTopology2019,Achouri2015a,Pfeiffer2013,Chen2020}, it will be outside the scope of this work.

Ideally, a designer would like to produce EMMS parameters that achieve desired far-field criteria such as beam width, side lobe level, or null locations directly. Essentially, they would supply a set of masks constraining the far-field pattern. However, current EMMS parameter design typically requires knowledge of the fields near the EMMS, which is only available for a few known transformations such as refraction, reflection, and collimation \cite{Selvanayagam2013, Budhu2020a}. For example, when performing refraction at $45^{\circ}$, a designer knows the incident excitation near field. The transmitted scattered near-field is specified to be a plane wave travelling $45^{\circ}$ away from the surface. The reflected scattered near field will be specified to be zero for a bianisotropic EMMS. With a complete set of near fields, one can use the generalized sheet transition conditions (GSTCs) to find the required EMMS surface parameters for these prescribed near fields \cite{EpsteinOptics2016}. For an arbitrary far-field, there is not a unique set of near fields or surface currents. As a result we cannot directly determine EMMS surface parameters in this fashion. In order to resolve this, there has been some research into designing EMMS parameters for more arbitrary far-field patterns.

Recently, the source reconstruction method (SRM) has been leveraged to synthesize EMMS parameters. It takes a desired far-field radiation pattern from the surface and aims to reconstruct an equivalent source to determine the required surface parameters. The SRM is used to derive surface impedance based on far-field criteria such as beam width, null location, and direction \cite{Brown2018,Brown2019,Brown2020,Brown2020a,Salucci2018}. 

There has also been work on the synthesis of reflectarray metasurfaces incorporating mutual coupling and edge effects \cite{Lang2018,Budhu2020a}. Lang \textit{et al} formulate the problem to maximize wireless power transfer to a far-field receiver by optimizing reflectarray element phases. The reflectarray elements incorporate mutual coupling and edge effects between them to get a more accurate model. In doing so, they are able to surpass typical reflectarray design methods to maximize gain in a given direction. Budhu \textit{et al} use the method of moments (MoM) to more accurately model a 3-layer reflectarray for collimating an incident cylindrical wave \cite{Budhu2020a}. By accounting for the mutual coupling and edge effects, typically ignored, they achieve good results (20.68 dB directivity for a 20$\lambda$ surface) with a faster, more efficient method. The main drawback is that it requires knowledge of the tangential fields above and below the surface. Another example utilizing the MoM with Fourier Bessel basis functions (FBBFs) was proposed for more arbitrary beam shaping \cite{Bodehou2019}. While this method does not rely on local periodicity assumptions, the FBBFs are entire domain basis functions that are only defined over an elliptical aperture. In addition, the algorithm requires a fully formed pattern to realize rather than far-field criteria directly. 

EMMS surface design has been approached using stochastic methods such as differential evolution algorithms and machine learning \cite{Rocca2009,Rocca2011, Li2018}. While both methods saw success, stochastic methods are non-deterministic. Synthesis using stochastic methods is often solution-specific because they require extensive algorithm parameter tuning for each instance. While all of these techniques are promising, they all suffer from important drawbacks.

In this work, we present a new method to determine the EMMS parameters, which satisfy far-field design goals. Starting with an EMMS model derived using the MoM, we are able to fully capture edge effects and mutual coupling between EMMS elements. We then use the alternating direction method of multipliers (ADMM) to deterministically optimize the EMMS parameters to satisfy far-field criteria including main beam level, side lobe level, and null location. In addition, we are also able to impose constraints on the curvature of the surface currents in order to allow for more manufacturable surfaces.

The paper is structured as follows. In \Cref{EMMS_Model}, we begin by discussing how the formulation of the EMMS model from the MoM. Following this, in \Cref{OptScheme}, we outline how the ADMM optimizer is constructed. Lastly, in \Cref{Examples}, we provide two-dimensional optimization examples demonstrating general multi-criteria optimization, extreme angle refraction with electrically small EMMSs, and Chebyshev beamforming before offering some concluding remarks.



\section{Electromagnetic Surface Model} \label{EMMS_Model}

\begin{figure}[!t]
\centering
\includegraphics[width=2.5in]{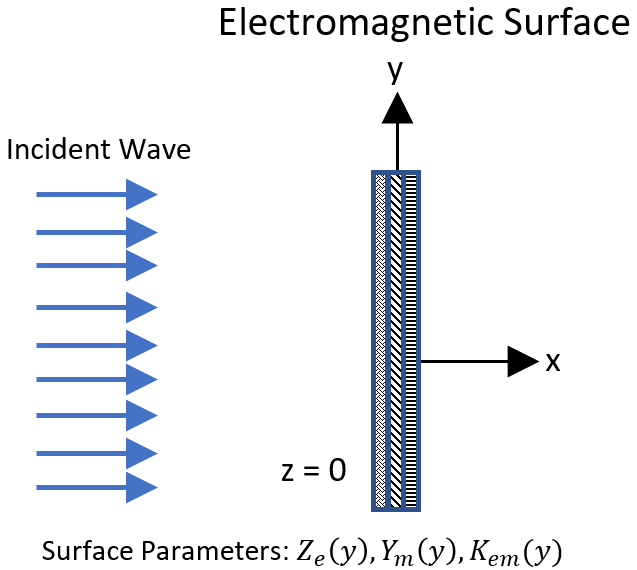}
% where an -eps-converted-to.pdf filename suffix will be assumed under latex, 
% and a .pdf suffix will be assumed for pdflatex; or what has been declared
% via \DeclareGraphicsExtensions.
\caption{Two dimensional EMMS configuration.}
\label{EMMSDiagram}
\end{figure}

Our two-dimensional model for the EMMS is constructed using the electric and magnetic field integral equations (EFIE and MFIE respectively), and solved using the MoM \cite{Gibson2015}. The EMMS is located on the $y$-axis centred about the origin. A simple diagram of its configuration is shown in \Cref{EMMSDiagram}. Throughout this paper we will consider the TM-polarized case ($\hat{z}$-directed E-field).  
%We begin with the bianisotropic sheet transition conditions (BSTCs) \cite{Epstein2016}.
%\begin{align}
%\overrightarrow{E}_{t,avg}=Z_{e}(y)\overrightarrow{J}_{s}(y)-K_{em}(y)[\hat{z}\times\overrightarrow{M}_{s}(y)]\\
%\overrightarrow{H}_{t,avg}=Y_{m}(y)\overrightarrow{J}_{s}(y)-K_{em}(y)[\hat{z}\times\overrightarrow{J}_{s}(y)]
%\end{align}
%(\textbf{Note to Sean:} in Ariel's BSTCs and elsewhere they use the average field rather than the total field that we use. Is the difference just a factor of 0.5 or are they equivalent?) We start with the electric field integral equation (EFIE) which can be used to find the induced surface currents. 
We first discretize the far-field into $M$ angular samples and the surface into $N$ spatial samples along the surface. We then apply the discretized EFIE for the more general bianisotropic EMMS case, which can be written as
\begin{align} \label{EFIE}
\overline{E}^{inc}_{tan}+\overline{E}^{scat}_{tan} = \overline{\overline{Z}}_{se}\overline{J}_z(y)-\overline{\overline{K}}_{em}\overline{M}_y(y)
\end{align}
where the electric and magnetic surface currents $\overline{J}_z(y) \in \mathbb{C}^{N}$, $\overline{M}_y(y) \in \mathbb{C}^{N}$ are $\hat{z}$-directed and $\hat{y}$-directed respectively, while $\overline{\overline{Z}}_{se} \in \mathbb{C}^{N\times N}$ and $\overline{\overline{K}}_{em} \in \mathbb{C}^{N\times N}$ are diagonal matrices containing the surface impedance and magneto-electric coupling coefficient as a function of position. The magnetic field integral equation (MFIE) can be constructed in a similar way and written as
\begin{align}\label{MFIE}
\overline{H}^{inc}_{tan}+\overline{H}^{scat}_{tan} = \overline{\overline{Y}}_{sm}\overline{M}_y(y)+\overline{\overline{K}}_{em}\overline{J}_z(y).
\end{align}
where $\overline{\overline{Y}}_{sm} \in \mathbb{C}^{N\times N}$ is a diagonal matrix containing the surface admittance.

The scattered electric field in \eqref{EFIE} from a thin strip along the $y$-axis is given by
\begin{align}\label{Escat}
\overline{E}_{scat}(\rho) 	= & -\frac{\omega\mu_0}{4}\int_{-w/2}^{w/2}\overline{J}_{z}(y')H_0^{(2)}(k|y-y'|)dy' \hat{z}
\end{align}
where $w$ is the width of the thin strip. To discretize this, we expand the current $\overline{J}_z(y)$ with pulse basis functions as
\begin{subequations}
\begin{align}
\overline{J}_{z}(y) &= \sum_{n=1}^N I^e_n f_n(y)\\
f_n(y) &= 
\begin{cases} 
0 & y < (n-1)\Delta_y \\
1 & (n-1)\Delta_y \leq y \leq n\Delta_y  \\
0 & n\Delta_y  < y
\end{cases}
\end{align}
\end{subequations}
which yields the matrix equation
\begin{subequations}
\begin{align}
\textit{\textbf{E}}_{scat} &= \textbf{Z}_e\textit{\textbf{I}}^e\\
\textbf{Z}_e(u,v) &= \frac{\omega\mu_0}{4}\int_{(v-1)\Delta y}^{v\Delta y} H_0^{(2)}(k |y(u)-y'|)dy'.
\end{align}
\end{subequations}

We can similarly derive the scattered magnetic field for \eqref{MFIE} as
\begin{align}\label{Hscat}
\begin{split}
\overline{H}_{scat}(\rho) 	=& -\frac{1}{4\omega\mu_0}(k^2+\frac{\partial^2}{\partial y^2})\\ &\int_{-w/2}^{w/2}\overline{M}_{y}(y')H_0^{(2)}(k|y-y'|)dy' \hat{z}.
\end{split}
\end{align}
and subsequently discretize it. Using pulse basis functions as before to represent $\overline{M}_y(y)$
\begin{subequations}
\begin{align}
\textit{\textbf{H}}_{scat} =& \textbf{Z}_m\textit{\textbf{I}}^m\\
\begin{split}
\textbf{Z}_m(u,v) =& \frac{k^2}{8\omega\mu_0}\int_{(v-1)\Delta y}^{v\Delta y} H_2^{(2)}(k |y(u)-y'|)  \\&+ H_0^{(2)}(k |y(u)-y'|)dy'
\end{split}
\end{align}
\end{subequations}

Using point matching we can construct two MoM equations, which capture the mutual coupling and edge effects of an EMMS. The resulting system of equations is
\begin{subequations}
\begin{align}
\textit{\textbf{E}}^{inc} &= \textbf{Z}_e\textit{\textbf{I}}^e+\textbf{Z}_{se}\textit{\textbf{I}}^e-\textbf{K}_{em}\textit{\textbf{I}}^m\\
\textit{\textbf{H}}^{inc} &= \textbf{Z}_m\textit{\textbf{I}}^m+\textbf{Y}_{sm}\textit{\textbf{I}}^m+\textbf{K}_{em}\textit{\textbf{I}}^e.
\end{align}
\end{subequations}
In our examples we would like to consider purely passive and lossless EMMSs because they are more efficient and easier to realise. As a result, $\textbf{Z}_{se}$ and $\textbf{Y}_{sm}$ will be purely imaginary, while $\textbf{K}_{em}$ is purely real \cite{Epstein2016}. This yields the equations
\begin{subequations} \label{PLMoMEqns}
\begin{align}
\textit{\textbf{E}}^{inc} &= \textbf{Z}_e\textit{\textbf{I}}^e+j\textbf{X}_{se}\textit{\textbf{I}}^e-\textbf{K}_{em}\textit{\textbf{I}}^m\\
\textit{\textbf{H}}^{inc} &= \textbf{Z}_m\textit{\textbf{I}}^m+j\textbf{B}_{sm}\textit{\textbf{I}}^m+\textbf{K}_{em}\textit{\textbf{I}}^e.
\end{align}
\end{subequations}

We also require matrix equations describing the far-field scattering from the induced surface currents on the EMMS. Using the two dimensional free space dyadic Green's function we can describe the far-field electric field due to the electric at a distance $\rho$ as
\begin{align}\label{Eff_E}
\overline{E}^{ff}_e(\phi) = -\frac{\omega\mu_0}{4} \sqrt{\frac{2}{\pi k \rho}}e^{j(\frac{\pi}{4}-k\rho)} \int_{-w/2}^{w/2}\overline{J}_{z}(y')e^{jky'\sin\phi}dy'.
\end{align}
We can likewise find the far-field electric field due to magnetic currents as
\begin{align}\label{Eff_H}
\begin{split}
\overline{E}^{ff}_m(\phi) =& -\frac{\omega\varepsilon_0 \eta_0}{4} \sqrt{\frac{2}{\pi k \rho}}e^{j(\frac{\pi}{4}-k\rho)} \\ &\int_{-w/2}^{w/2}\overline{M}_{y}(y')e^{jky'\sin\phi}\cos \phi dy'.
\end{split}
\end{align}

We can again discretize the current using pulse basis functions to yield the matrix equations
\begin{subequations}\label{GIMatrix}
\begin{align}
\textit{\textbf{E}}_{e}^{ff} = & \textbf{G}^e\textit{\textbf{I}}^e\\
\textbf{G}^e(u,v) = & -\frac{\omega\mu0}{4} \sqrt{\frac{2}{k\pi}}e^{j\frac{\pi}{4}}\frac{e^{-jk\rho}}{\sqrt{\rho}}e^{jky(v)\sin(\phi(u))}\Delta y \\
\textit{\textbf{E}}_{m}^{ff} = & \textbf{G}^m\textit{\textbf{I}}^m\\
\textbf{G}^m(u,v) = & -\frac{\omega\epsilon_0\eta_0}{4} \sqrt{\frac{2}{k\pi}}e^{j\frac{\pi}{4}}\frac{e^{-jk\rho}}{\sqrt{\rho}}e^{jky(v)\sin(\phi(u))}\cos(\phi(u))\Delta y 
\end{align}
\end{subequations}
where $\textbf{G}^e \in \mathbb{C}^{M\times N}$ and $\textbf{G}^m \in \mathbb{C}^{M\times N}$ are matrices transforming the electric and magnetic surface currents to far-field electric field.

Using these equations describing the EMMS and its far-field scattered field, we can now assemble an optimization scheme around them. It is worth noting that we have chosen a simple example for a demonstration of this method. MoM can be expanded for other geometries in two and three dimensions. In addition, any feed interactions or multiple EMMSs could also be incorporated into this model. These augmentations would only require altering the impedance matrix $\textbf{Z}_e$ and $\textbf{Z}_m$ to account for the altered coupling between surface currents and calculating $\textbf{\textit{E}}^{inc}$ and $\textbf{\textit{H}}^{inc}$ for different excitation fields in \eqref{PLMoMEqns}.

\section{Optimization Scheme} \label{OptScheme}
\subsection{Convex Optimization Background}
Convex optimization is the process of minimizing a convex objective function subject to a set of convex constraints. A function $f(x)$ is deemed convex if
\begin{align}
f(\theta x + (1-\theta)y) & \leq \theta f(x) + (1-\theta)f(y) 
\end{align}
for all $x,y \in \Re$ and $\theta \in [0,1]$ \cite{Boyd2004}. A set $S$ is deemed convex if for any $x,y \in S$
\begin{align}
\theta x + (1-\theta) y  \in S
\end{align}
for $\theta \in [0,1]$ \cite{Boyd2004}. The general form of a convex problem is
\begin{equation}
\begin{aligned}
\minimize_x &\quad f_0(x)\\
\textrm{subject to} & \quad f_i(x) \leq 0 \quad i = 1,...,n \\
& \quad g_k(x) = 0 \quad k = 1,...,m
\end{aligned}
\end{equation}
where $f_0$ is a convex function called the objective function, $f_i$ are convex functions forming the set of inequality constraints, and $g_k$ are affine functions forming a set of equality constraints. A point $x_0$ is called feasible if $f_i(x_0) \leq 0 $ for $ i = 1,...,n$ and $g_k(x_0) = 0$ for $k = 1,..,m$. A point $x^*$ is deemed optimal if it is feasible and $f_0(x^*) < f_0(x_0)$ for all feasible points $x_0$.

The main challenge of convex optimization is formulating a given problem as a convex one. Once that is achieved, there exist many reliable solvers which can solve most convex problems in polynomial time \cite{CVX, Boyd2004}. 

\subsection{Alternating Direction Method of Multipliers Relaxation}

\begin{figure*}[!t]
\centering
\includegraphics[height=4in]{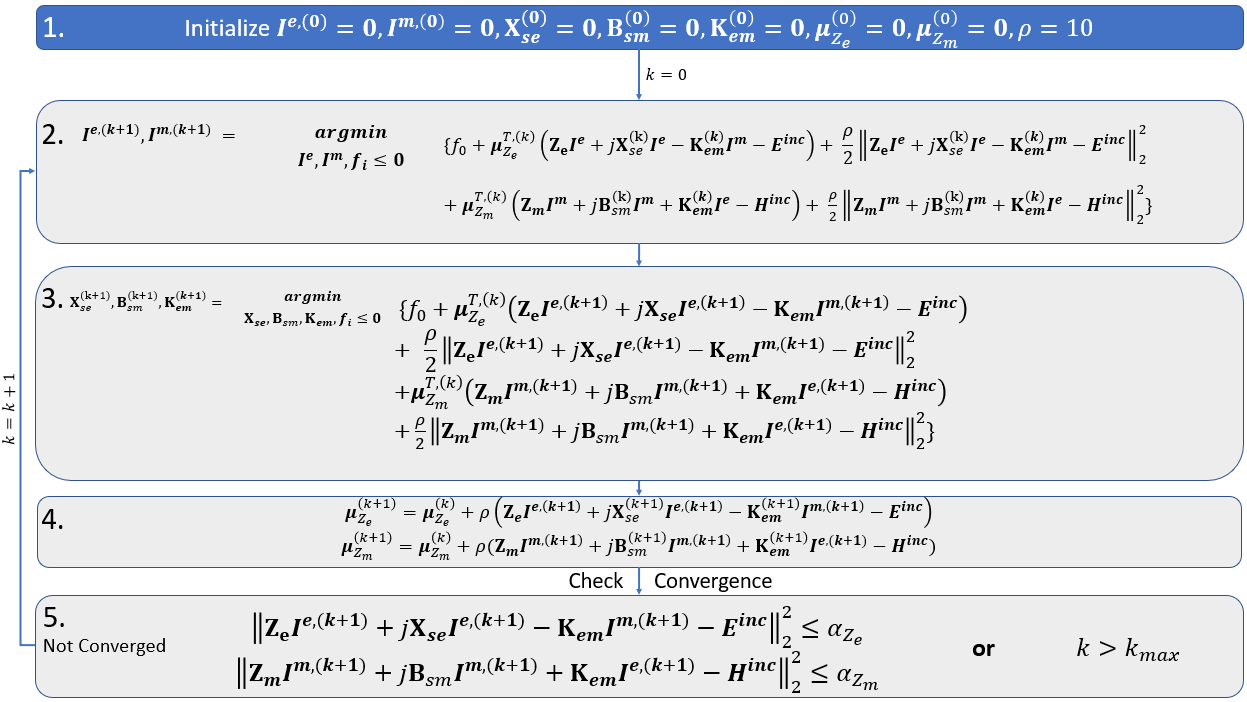}
% where an -eps-converted-to.pdf filename suffix will be assumed under latex, 
% and a .pdf suffix will be assumed for pdflatex; or what has been declared
% via \DeclareGraphicsExtensions.
\caption{Alternating direction method of multipliers steps: \textbf{1.} Initialize variables and set $\rho$ value. \textbf{2.} Minimize the augmented Lagrangian with respect to the electric and magnetic currents. \textbf{3.} Using the currents from step 2, minimize the augmented Lagrangian with respect to the surface parameters. \textbf{4.} Update the dual variables to reflect the amount of equality constraint violation. \textbf{5.} Check for convergence. Here we evaluate the amount of equality constraint violation as our convergence criteria. If it is less than $\alpha$ then terminate ADMM or a maximum number of iterations $k_{max}$, otherwise increment k and return to step 2.}
\label{ADMM_Diagram}
\end{figure*}

In order to optimize an EMMS, we need to formulate the problem as a convex one. This has a form similar to
\begin{subequations} \label{EMMS_basic_form}
\begin{align}
\minimize_{\textit{\textbf{I}}^e,\textit{\textbf{I}}^m,\textbf{X}_{se},\textbf{B}_{sm},\textbf{K}_{em}} &\quad f_0(\textit{\textbf{I}}^e,\textit{\textbf{I}}^m,\textbf{X}_{se},\textbf{B}_{sm},\textbf{K}_{em})\\
\textrm{subject to} & \quad \textit{\textbf{E}}^{inc} = \textbf{Z}_e\textit{\textbf{I}}^e+j\textbf{X}_{se}\textit{\textbf{I}}^e-\textbf{K}_{em}\textit{\textbf{I}}^m \label{EMMS_E_const}\\
& \quad \textit{\textbf{H}}^{inc} = \textbf{Z}_m\textit{\textbf{I}}^m+j\textbf{B}_{sm}\textit{\textbf{I}}^m+\textbf{K}_{em}\textit{\textbf{I}}^e. \label{EMMS_H_const}\\
& \quad f_i(\textit{\textbf{I}}^e,\textit{\textbf{I}}^m,\textbf{X}_{se},\textbf{B}_{sm},\textbf{K}_{em}) \leq 0 \; i = 1,...,n 
\end{align}
\end{subequations}
where $f_0$ and $f_i$ are some objective and constraint functions for a certain design goal. The optimization variables in this case are the surface current coefficients $\textit{\textbf{I}}^e$ and $\textit{\textbf{I}}^m$ along with the surface reactance ($\textbf{X}_{se}$), susceptance ($\textbf{B}_{sm}$), and magneto-electric coupling ($\textbf{K}_{em}$). The advantage of this formulation is that we are able to optimize directly for passive and lossless surface parameters while incorporating all of the physics of the problem. Solving directly for the currents instead would then require another step to convert to a passive and lossless solution with the GSTCs. Solving for currents directly has been employed by other authors, but often requires global optimizers \cite{Brown2020a}. 

Unfortunately, the $j\textbf{X}_{se}\textit{\textbf{I}}^e-\textbf{K}_{em}\textit{\textbf{I}}^m$ and $j\textbf{B}_{sm}\textit{\textbf{I}}^m+\textbf{K}_{em}\textit{\textbf{I}}^e$ terms in \eqref{EMMS_E_const} and \eqref{EMMS_H_const} are non-convex. This is because they are bi-affine (two variables multiplied together). When confronted with a non-convexity there are two options. One can reformulate the model to eliminate the non-convex terms or relax the non-convexity. Reformulating the model without biaffine terms does not seem to be possible in our case. Therefore, we must relax the problem.

A relaxation typically allows a non-convex problem to be formulate as a convex one at the expense of an enlarged feasibility region. As a result, it is important to ensure any solutions provided by the optimizer are also feasible for our original problem. This seems like a heavy price to pay. However, many relaxations provide solutions which are tight to the original problem. There exist many relaxations for biaffine problems \cite{Sherali1992,SheraliRLT1992}. We will use ADMM \cite{Boyd2011}.

ADMM first forms the augmented Lagrangian and then alternatingly minimizes the augmented Lagrangian with respect to one of the biaffine variables each iteration. Following this, it updates the dual variables to penalize equality constraint violation. Doing so, it arrives at a relaxed solution to the previously unsolvable problem. Convergence for ADMM in this case is confirmed by ensuring that the equality constraint violation is lower than a certain threshold $\alpha_{Z_e}$ and $\alpha_{Z_m}$. A flow chart in \Cref{ADMM_Diagram} shows the steps of the algorithm as implemented for the EMMS. 

ADMM can be performed using the CVX Matlab package \cite{CVX}. CVX is a free tool which allows convex optimization problems to be input in a fashion similar to \eqref{EMMS_basic_form} and then solved. The convex optimization solver that we use is Mosek, which is available on the CVX academic license. 

One thing to note is that complex terms should be separated into their real and imaginary components. For example one can decompose a vector $\textbf{a}$ and matrix $\textbf{A}$ as
\begin{align*}
\textbf{a} \rightarrow
\begin{bmatrix}
\textrm{Re}\lbrace \textbf{a}\rbrace \\
\textrm{Im}\lbrace \textbf{a} \rbrace
\end{bmatrix}
and \;
\textbf{A}\rightarrow
\begin{bmatrix}
\textrm{Re}\lbrace \textbf{A} \rbrace & -\textrm{Im}\lbrace \textbf{A}\rbrace \\
\textrm{Im}\lbrace \textbf{A} \rbrace & \textrm{Re}\lbrace \textbf{A} \rbrace
\end{bmatrix}.
\end{align*}
This is because most convex optimization solvers can't work directly with complex numbers. For the rest of the paper we will assume that complex terms will be converted into this form before being solved. It is worth noting no problem information is lost in this conversion.

\subsection{Electromagnetic Metasurface Surface Parameter Optimizer}
With ADMM, we have a way to relax and subsequently optimize the surface currents and parameters within the physical limits of the EMMS. The remaining step is to construct constraints and/or objective functions corresponding to potential design specifications. 
\subsubsection{Main Beam Level}\label{MBLevel_section}
To try to force a beam to achieve a certain level (or get as close as possible) we can add a $\ell_2$-norm minimization term to the objective. For example, to force a beam at $\phi_0$ to a level $MB_{level}$ the objective function can take the form
\begin{align}\label{MBLevel}
\begin{split}
f_{MB}(\phi_0) =& \Vert \textbf{G}^e(\phi_0)\textbf{\textit{I}}^e+\textbf{G}^m(\phi_0)\textbf{\textit{I}}^m \\&+ \textbf{E}_{inc}^{ff}(\phi_0) - MB_{level}\Vert_2^2,
\end{split}
\end{align}
where $\textbf{E}_{inc}^{ff}(\phi_0)$ is the incident field in the spectral domain. This is required for the total field because $\textbf{G}^e(\phi_0)\textbf{\textit{I}}^e+\textbf{G}^m(\phi_0)\textbf{\textit{I}}^m$ represents only the scattered field. One can form multiple main beams by supplying a range of angles to $\phi_0$, leading to more rows of \eqref{MBLevel}.

\subsubsection{Null Location} \label{NullLocation_section}
To form a null at $\phi_n$ in the far-field we can add the term,
\begin{align}\label{Null}
f_{null}(\phi_n)= \Vert \textbf{G}^e(\phi_n)\textbf{\textit{I}}^e+\textbf{G}^m(\phi_n)\textbf{\textit{I}}^m + \textbf{E}_{inc}^{ff}(\phi_n)\Vert_2^2
\end{align}
to the objective function. This is similar to \eqref{MBLevel} but the level to achieve is $0 \; \textrm{V/m}$. Similar to \Cref{MBLevel_section}, one can require multiple null angles by supplying a range of angles to $\phi_n$. 

\subsubsection{Maximum Side Lobe Level} \label{SLL_section}
Enforcing a maximum permissible side lobe over a certain set of angles can be done with the inequality constraint
\begin{align}
\vert \textbf{G}^e(SL)\textbf{\textit{I}}^e+\textbf{G}^m(SL)\textbf{\textit{I}}^m + \textbf{E}_{inc}^{ff}(SL) \vert \leq \tau + slack_{SL},
\end{align}
where we force the absolute value of the total field over the side lobe region $SL$ to be less than a certain side lobe level $\tau$. We've included a slack variable $slack_{SL}$ so the following term must be added to the objective:
\begin{align}
f_{SL} = \Vert slack_{SL} \Vert^2_2
\end{align}
This is required in order to make the inequality active. The reason the slack variable is included is so that the optimizer has some ``room" during the preliminary iterations to balance other parts of the optimization. Without it, ADMM would struggle to make progress for more challenging problems. This is further explored in \Cref{Examples}.

\subsubsection{Surface Current Smoothness} \label{CurrentSmooth_section}
A useful term to add to the optimizer is a constraint on the limit of second derivative allowed by the surface currents $\textbf{\textit{I}}^e$ and $\textbf{\textit{I}}^m$. This can be done with the inequality constraints,
\begin{align}
\vert \textbf{D}\textbf{\textit{I}}^e\vert \leq D_{max}^e + slack_{D^e}\\
\vert \textbf{D}\textbf{\textit{I}}^m\vert \leq D_{max}^m + slack_{D^m}
\end{align}
where $\textbf{D}$ is the discrete second derivative matrix described by,
\begin{align}
\frac{1}{{\Delta y}^2}
\begin{bmatrix}
1 & 2 & 1 & 0 & \cdots & \\
0 & 1 & 2 & 1 & 0 & \cdots \\
 &  & \ddots & \ddots & \ddots \\
0 & & \cdots & 1 & 2 & 1
\end{bmatrix}
\end{align} 
and $\Delta y = W/N$ is the distance between samples along the surface. Note that $\textbf{D} \in \Re^{(N-2)\times N}$ as the the endpoints are allowed to be non-zero. As before the slack variables $slack_{D^e}$ and $slack_{D^m}$ must be minimized in the cost function with,
\begin{align}
f_{D} = \Vert slack_{D^e} \Vert^2_2 + \Vert slack_{D^m} \Vert^2_2
\end{align}
The advantage of constraining the curvature of the currents is two-fold. Firstly, it makes the EMMSs easier to realize. This is because it is much easier to design unit cells if there is not a very large jump in surface currents from cell to cell. Secondly, because we use pulse basis functions, if the currents are allowed to be highly erratic, it could lead to some non-physical results. Similar current regularizations have been used by other authors as well \cite{Salucci2019}.

\begin{figure*}[!h]
\begin{subequations} \label{EMMS_total_form}
\begin{align}
\minimize_{\substack{\textit{\textbf{I}}^e,\textit{\textbf{I}}^m,\textbf{X}_{se},\textbf{B}_{sm},\textbf{K}_{em},slack_{SL},slack_{D^e},slack_{D^m}}} &\quad \alpha_{MB}f_{MB}(MB)+\alpha_{NULL}f_{null}(NU)+\alpha_{SL}f_{SL}+\alpha_{D}f_{D}\\
\textrm{subject to} \qquad \qquad \ \  & \quad \textit{\textbf{E}}^{inc} = \textbf{Z}_e\textit{\textbf{I}}^e+j\textbf{X}_{se}\textit{\textbf{I}}^e-\textbf{K}_{em}\textit{\textbf{I}}^m \\
& \quad \beta(\textit{\textbf{H}}^{inc} = \textbf{Z}_m\textit{\textbf{I}}^m+j\textbf{B}_{sm}\textit{\textbf{I}}^m+\textbf{K}_{em}\textit{\textbf{I}}^e) \label{H_eqn}\\
& \quad\vert \textbf{G}^e(SL)\textbf{\textit{I}}^e+\textbf{G}^m(SL)\textbf{\textit{I}}^m + \textbf{E}_{inc}^{ff}(SL) \vert \leq \tau + slack_{SL} \\
& \quad \vert \textbf{D}\textbf{\textit{I}}^e\vert \leq D_{max}^e + slack_{D^e} \\
& \quad \vert \textbf{D}\textbf{\textit{I}}^m\vert \leq D_{max}^m + slack_{D^m}
\end{align}
\end{subequations}
\hrulefill
\end{figure*}

\subsubsection{Complete Formulation}
We can assemble a complete formulation if we fill in the objective function and inequality constraints of \eqref{EMMS_basic_form} with \Cref{MBLevel_section,NullLocation_section,SLL_section,CurrentSmooth_section}. This can be written as in \eqref{EMMS_total_form}, where $MB \in [0^\circ, 360^{\circ}]$ represents the set of main beam angles, $NU \in [0^\circ, 360^{\circ}]$ represents the set of null angles, $SL \in [0^\circ, 360^{\circ}]$ is the set of angles comprising the side lobe region, and $\alpha_{MB},\alpha_{NULL},\alpha_{SL},\alpha_{D}$ are predetermined weights for different terms in the objective function. Due to different magnitudes of the electric and magnetic current MoM equations, a scaling term is needed. The $\beta$ term in is \eqref{H_eqn} used to scale the magnetic current MoM equation. This is because when the augmented Lagrangian is formed, the equality constraints compete for minimization. Typical $\beta$ values are 1000. This was determined empirically. The weights are used if different portions of the optimization should be stressed more. For example, if the optimizer is having trouble satisfying the side lobe level constraint, $\alpha_SL$ could be increased relative to the other weights. This is done on an experimental basis. 

\section{Examples} \label{Examples}
It is worth noting that the supplied design goals must be first and foremost physically feasible for the EMMS. While the success of the optimization tool could be used as a rough guide for far-field pattern feasibility, it is not rigorous.

In the following examples we choose to terminate ADMM after a certain number of iterations (usually 150-200). This offers a good trade-off between far-field criteria and equality constraint satisfaction. For large $15\lambda$ EMMSs, 150 iterations takes roughly an hour on a typical desktop PC running CVX on Matlab for $15\lambda$ surfaces with $N=501$ spatial and $M=361$ angular samples. 

\subsection{Multi-criteria Example} \label{MultiCrit}
In order to demonstrate this methodology we will first present a multi-criteria optimization example. This example will combine all of the criteria in \Cref{MBLevel_section,NullLocation_section,SLL_section,CurrentSmooth_section}. 

An example of far-field parameters to optimize can be seen in \Cref{MultiCrit_table}. Although they are not derived from a concrete antenna engineering scenario, they were chosen as a diverse set of constraints to demonstrate the capabilities of this method. The results of the optimization for a bianisotropic EMMS can be seen in \Cref{MultiCrit}. For this example, we don't make use of the side lobe slack variable because the optimizer does not seem to have trouble meeting the side lobe demands.

\begin{table}[!t]
\caption{Multi criteria optimization values}
\label{MultiCrit_table}
\centering
\begin{tabular}{|c||c|}
\hline
 & \bfseries Value \\
\hline
\bfseries Surface Width ($W$) & $15\lambda$ \\
\hline
\bfseries Main Lobe Angles  ($\phi_0$) & $\left\lbrace 45^{\circ},130^{\circ},315^{\circ}\right\rbrace$ \\
\hline
\bfseries Main Lobe Level ($MB_{level}$) & 6.36 \\
\hline
\bfseries Side Lobe  & $\lbrace 0^{\circ}:35^{\circ},90^{\circ}:120^{\circ},$ \\
\bfseries Angles ($SL$) &  $ 140^{\circ}:270^{\circ},325^{\circ}:360^{\circ}\rbrace$\\
\hline
\bfseries Side Lobe Level ($\tau$)& 1.75\\
\hline
\bfseries Null Angles ($\phi_n$)& $\left\lbrace 80^{\circ},83^{\circ},85^{\circ}\right\rbrace$  \\
\hline
\bfseries $D^e_{max}$  & 0.1\\
\hline
\bfseries $D^m_{max} $ & 25\\
\hline
\bfseries $\left\lbrace \alpha_{SL},\alpha_{D}\right\rbrace$ &  $\left\lbrace 0,1\right\rbrace$ \\
\hline
\end{tabular}
\end{table}

\begin{figure}[!t]
\centering
\subfloat[]{\includegraphics[width=2.5in]{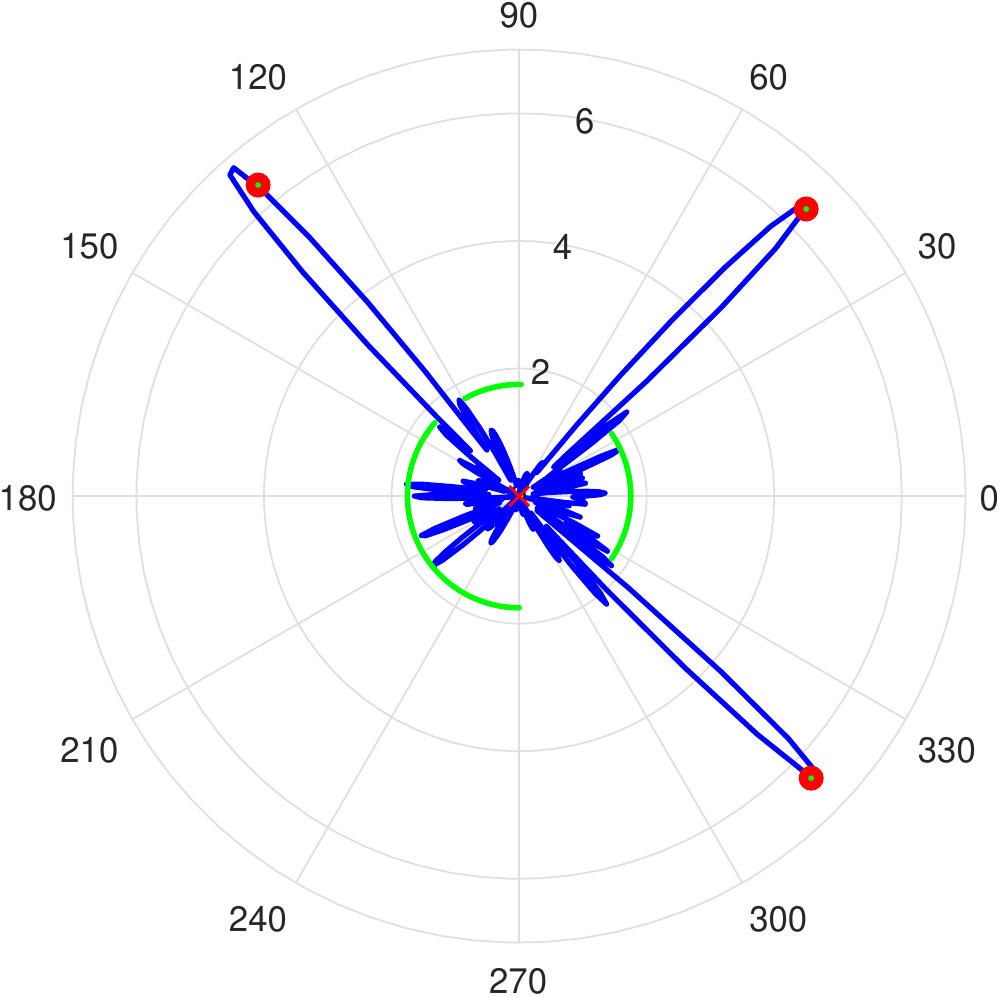}}\label{fig:MultiCritPolar}
\newline
\subfloat[]{\includegraphics[width=\textwidth/2]{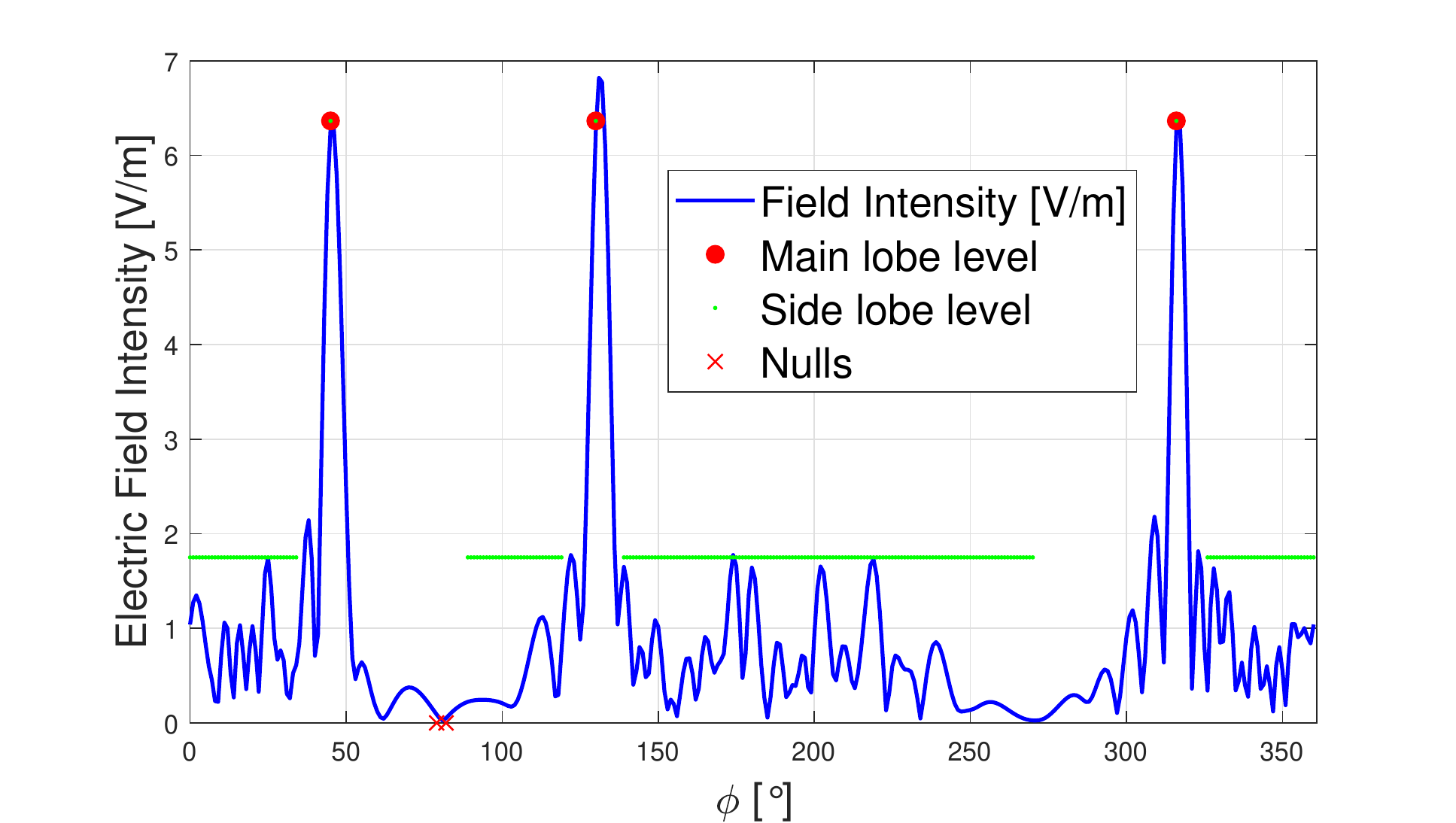}}\label{fig:MultiCritLinear}
% where an -eps-converted-to.pdf filename suffix will be assumed under latex, 
% and a .pdf suffix will be assumed for pdflatex; or what has been declared
% via \DeclareGraphicsExtensions.
\caption{Multi criteria optimization of far-field parameters with a bianisotropic EMMS. Total field in polar (a) and linear (b) format displayed for clarity.}
\label{MultiCrit}
\end{figure}

\begin{figure}[!t]
\centering
\includegraphics[width=3in]{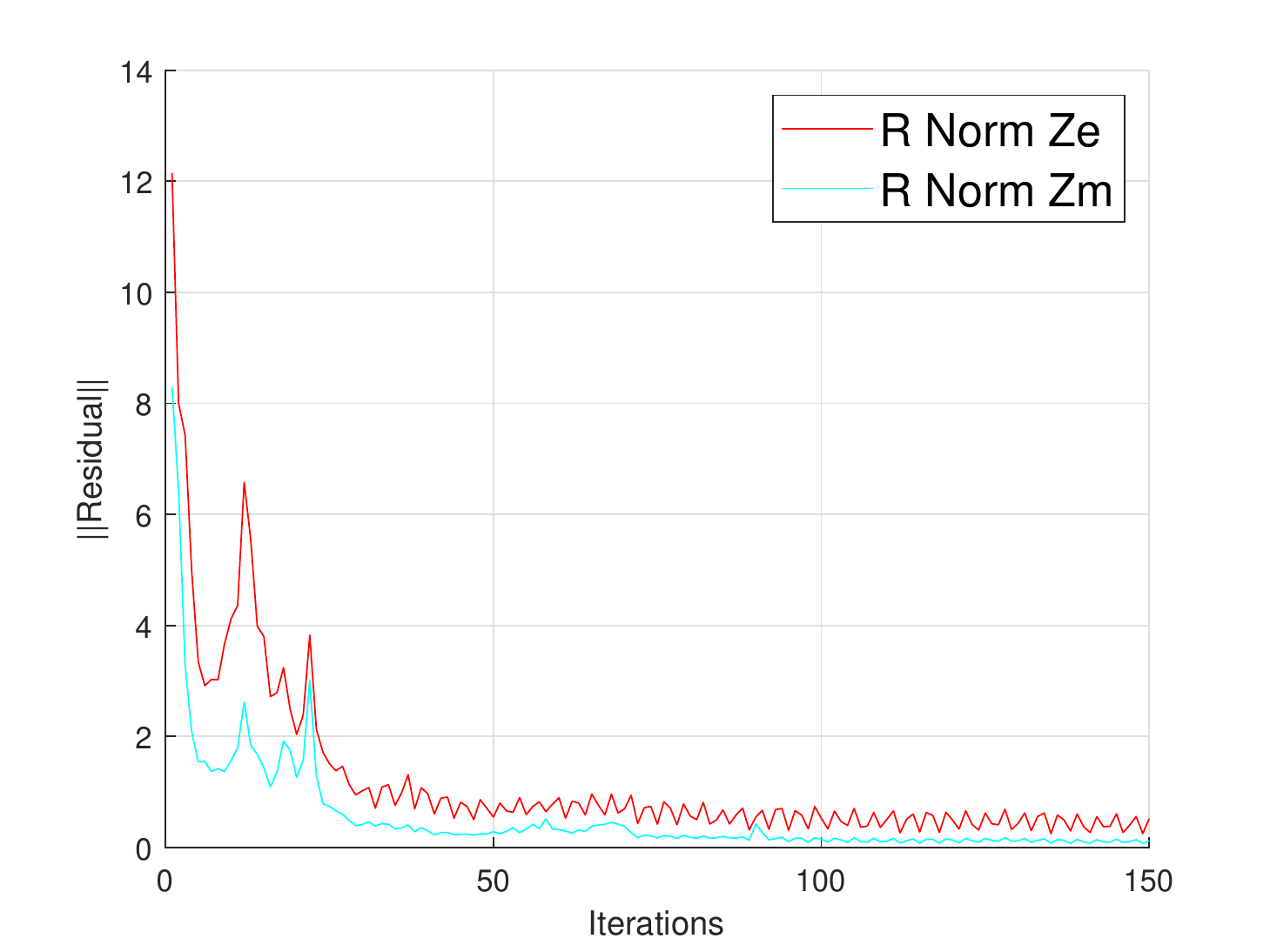}
% where an -eps-converted-to.pdf filename suffix will be assumed under latex, 
% and a .pdf suffix will be assumed for pdflatex; or what has been declared
% via \DeclareGraphicsExtensions.
\caption{Equality constraint residual for \Cref{MultiCrit}. Although it is not monotonically decreasing after 100 iterations it reaches empirically determined acceptable levels. }
\label{MultiCritRes}
\end{figure}

We can see the equality constraint residuals for each iteration in \Cref{MultiCritRes}. These residuals are the $\ell_2$-norm electric and magnetic MoM equality constraint violations $\Vert \textbf{Z}_e\textit{\textbf{I}}^e+j\textbf{X}_{se}\textit{\textbf{I}}^e-\textbf{K}_{em}\textit{\textbf{I}}^m - \textit{\textbf{E}}^{inc} \Vert^2_2$ and $\Vert  \textbf{Z}_m\textit{\textbf{I}}^m+j\textbf{B}_{sm}\textit{\textbf{I}}^m+\textbf{K}_{em}\textit{\textbf{I}}^e - \textit{\textbf{H}}^{inc} \Vert^2_2$ respectively. Although the residuals in \Cref{MultiCritRes} do not decrease monotonically, they do reduce quite effectively after only 100 iterations. This is to be expected because ADMM is not guaruanteed to converge monotonically with biaffine terms \cite{Boyd2011}. 

\begin{figure}[!t]
\centering
\includegraphics[width=\textwidth/2]{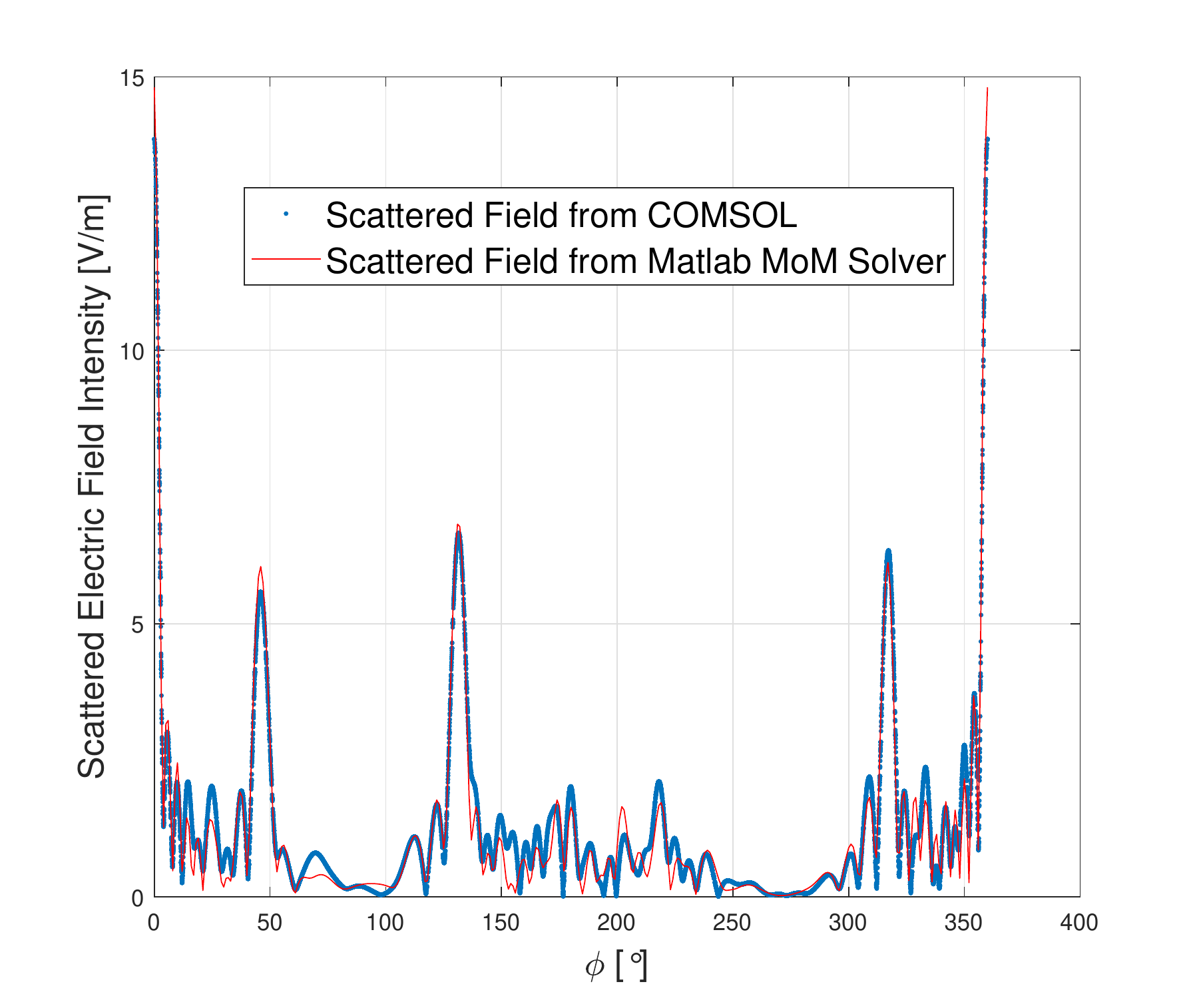}
% where an -eps-converted-to.pdf filename suffix will be assumed under latex, 
% and a .pdf suffix will be assumed for pdflatex; or what has been declared
% via \DeclareGraphicsExtensions.
\caption{Scattered electric field intensities comparison for multi-criteria optimization results with COMSOL \cite{COMSOL} and Matlab-based \cite{Matlab} MoM solver. As we can see the fields are fairly close so we can be confident of our results going forward.}
\label{MultiCritCOMSOL}
\end{figure}

In order to verify our results we input our surface parameters to a model in COMSOL, a commercial multiphysics solver. The scattered field can be seen in \Cref{MultiCritCOMSOL}. Note that the shadow of the surface is present in the scattered field but not the total field. The scattered field agrees well with our Matlab model so we can be confident in our results going forward. 
%\begin{figure}[!t]
%\centering
%\includegraphics[width=\textwidth/2]{images/MultiCrit_scalar-eps-converted-to.pdf}
%% where an -eps-converted-to.pdf filename suffix will be assumed under latex, 
%% and a .pdf suffix will be assumed for pdflatex; or what has been declared
%% via \DeclareGraphicsExtensions.
%\caption{Multi-criteria optimization of far-field parameters with a scalar EMMS.}
%\label{MultiCritScalar}
%\end{figure}

\subsection{Extreme Angle Refraction with Electrically Small EMMSs} \label{ExtremeRefrac}
\begin{figure}[!t]
\centering
\includegraphics[width=2.5in]{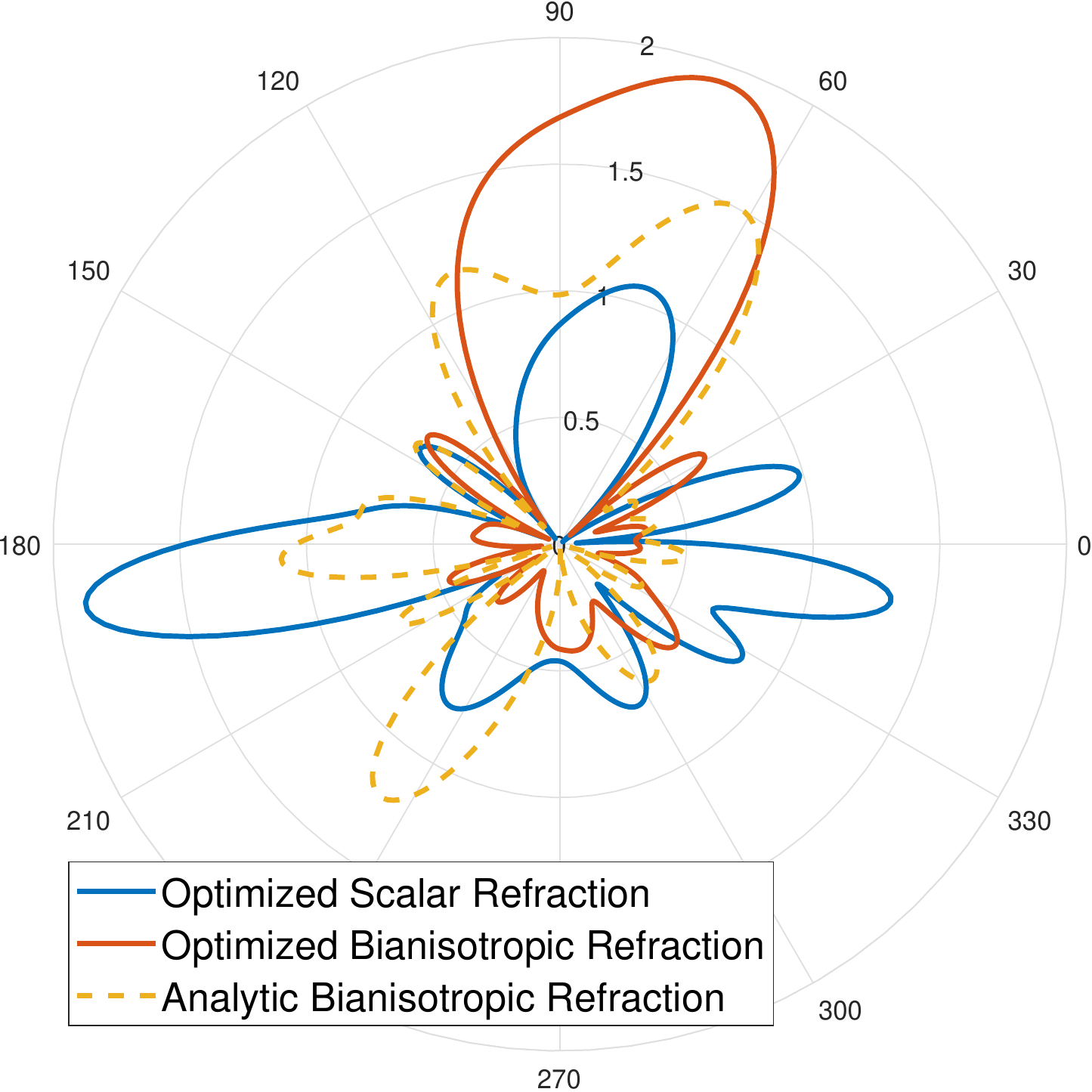}
% where an -eps-converted-to.pdf filename suffix will be assumed under latex, 
% and a .pdf suffix will be assumed for pdflatex; or what has been declared
% via \DeclareGraphicsExtensions.
\caption{Comparison of different methods of plane wave refraction to $72^{\circ}$ with an electrically small surface ($3\lambda$). Without the extra degree of freedom afforded by bianisotropy, the scalar optimized case exhibits a large reflected beam at $180^{\circ}$. We can see an improvement in beam direction and magnitude in comparison to the analytic formulation derived by Epstein \textit{et al} \cite{Epstein2016}.}
\label{SmallSurf}
\end{figure}

\begin{figure}[!t]
\centering
\includegraphics[width=2.5in]{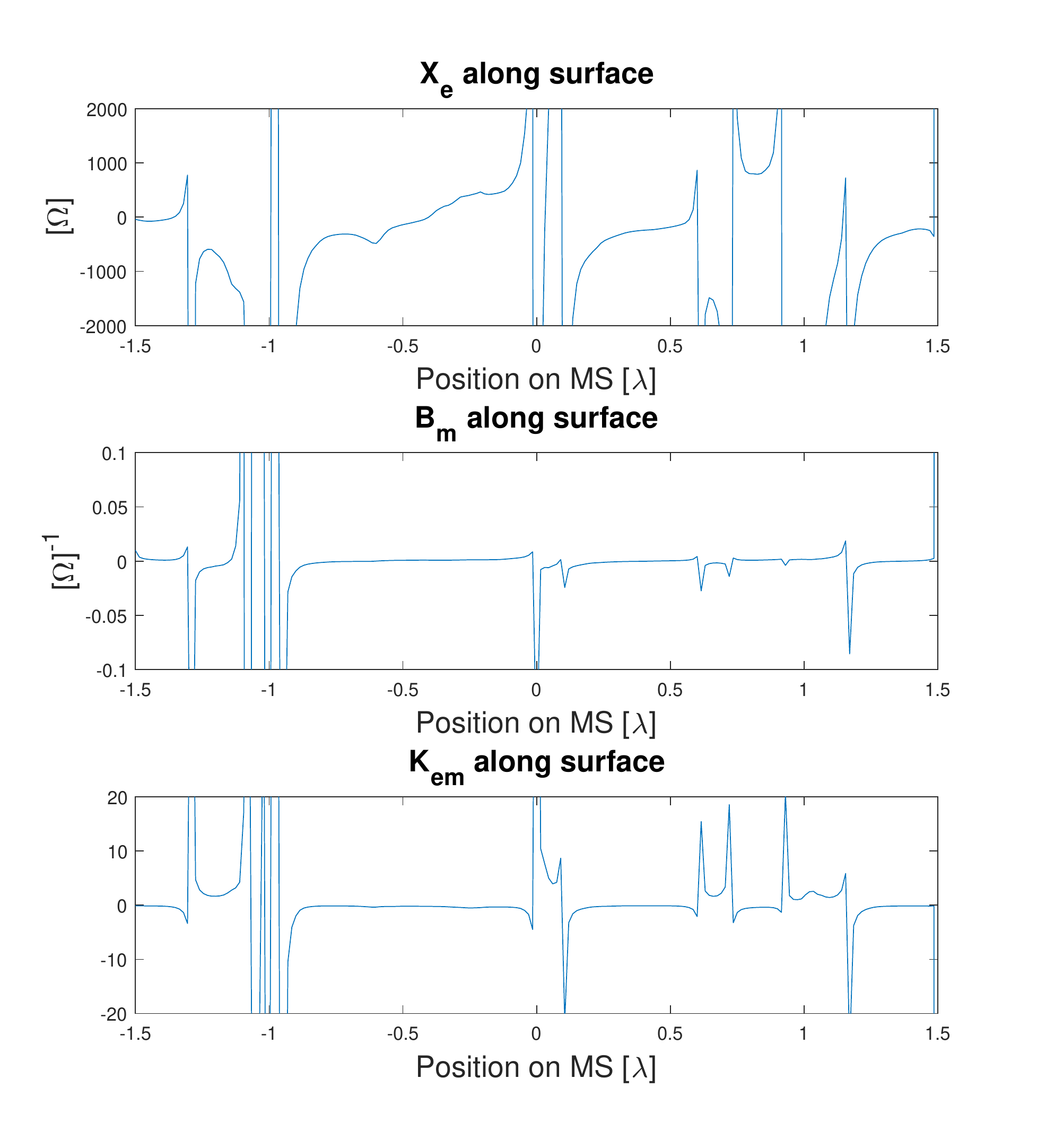}
% where an -eps-converted-to.pdf filename suffix will be assumed under latex, 
% and a .pdf suffix will be assumed for pdflatex; or what has been declared
% via \DeclareGraphicsExtensions.
\caption{Surface parameters for the optimized bianisotropic EMMS in \Cref{SmallSurf}}
\label{SmallSurfLoading}
\end{figure}

\begin{figure}[!t]
\centering
\includegraphics[width=2.5in]{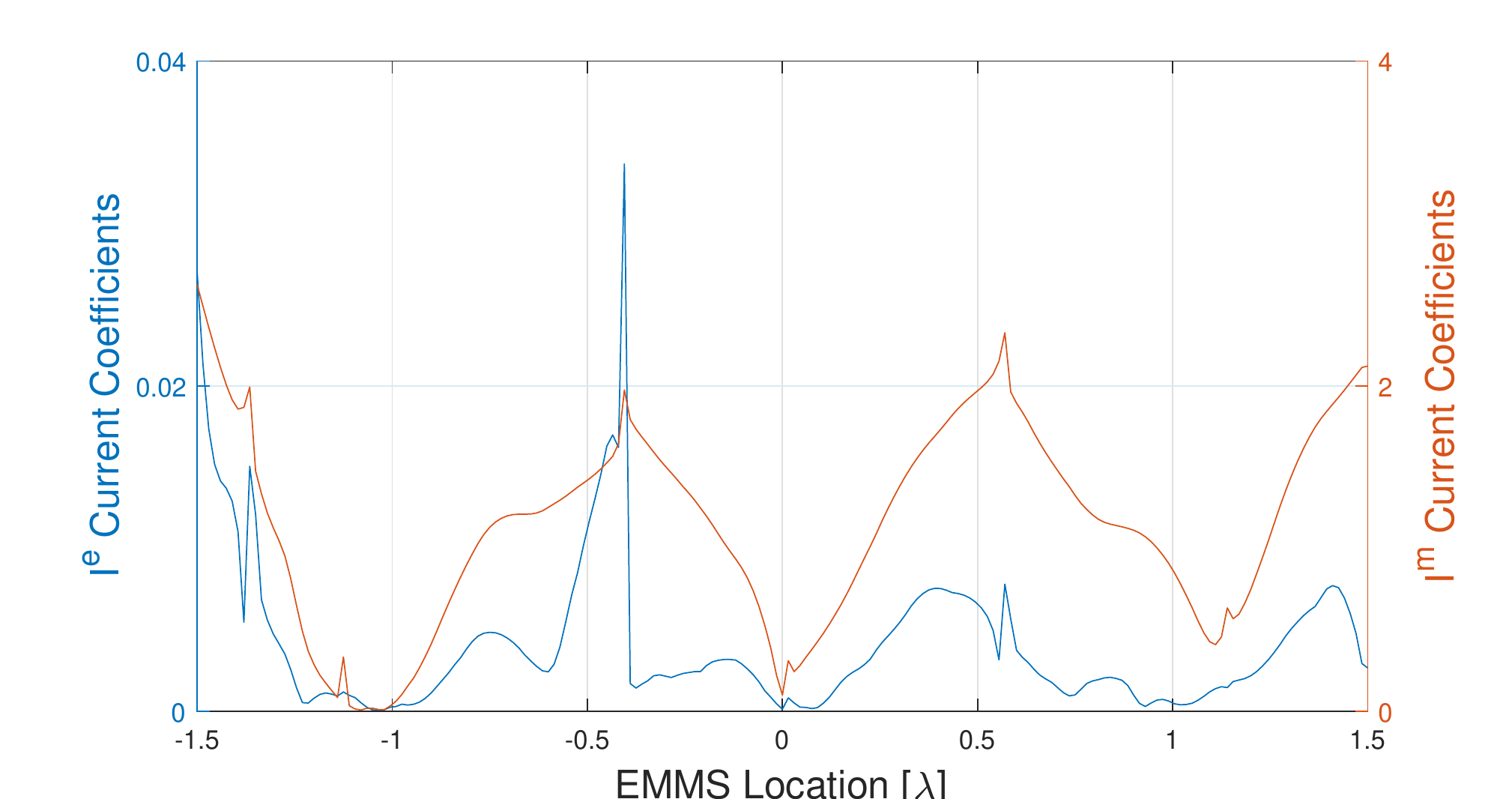}
% where an -eps-converted-to.pdf filename suffix will be assumed under latex, 
% and a .pdf suffix will be assumed for pdflatex; or what has been declared
% via \DeclareGraphicsExtensions.
\caption{Surface current coefficients for the optimized bianisotropic EMMS in \Cref{SmallSurf} }
\label{SmallSurfCurrents}
\end{figure}

Plane wave refraction is a well studied use case for EMMSs \cite{Epstein2016,EpsteinOptics2016,Selvanayagam2013}. These approaches typically begin with knowledge of the required near fields, which are determined analytically. With these desired near fields, they can then determine the surface parameters using the GSTCs. Although this is effective for larger EMMSs, this approach neglects edge effects. Edge effects play a big role for electrically small (for example $< 5 \lambda$) EMMSs. Our model, based off of the MoM, incorporates edge effects and mutual coupling between elements. As a result, we are able to perform plane wave refraction with these small EMMSs much more effectively. The MoM has been used to capture these effects before, but not for a transmissive surface \cite{Budhu2020}. We would also like to re-emphasize that we do not assume any knowledge of near or far field. We just supply desired characteristics, in this case a desired main beam level and direction, to the optimizer, which attempts to find the best surface parameters for this field transformation. The far-field results can be seen in \Cref{SmallSurf}, while the surface parameters and currents are displayed in \Cref{SmallSurfCurrents} and \Cref{SmallSurfLoading} respectively.

As we can see from the results, using the optimizer results in less beam pointing error than the analytic formulation derived by Epstein \textit{et al}\cite{Epstein2016}. This is mainly due to leveraging the edge effects incorporated in the MoM model. We can also see the difference in performance afforded by bianisotropy. There are much fewer reflections when allowing this extra degree of freedom.

\subsection{Chebyshev Beamforming} \label{Cheby}
An interesting far-field pattern to replicate is the Chebyshev array factor typically used to achieve a constant side lobe level. To the best of our knowledge, the EMMS surface parameters needed for this have never been solved for before. The far-field pattern we will use is the Chebyshev array factor, but because we have EMMS elements, rather than isotropic elements, we will use a very small uniform aperture as the element factor. As a result, we will have the typical Cheybshev array factor on the transmitted side with no reflections. 

\begin{figure}[!t]
\centering
\includegraphics[width=3in]{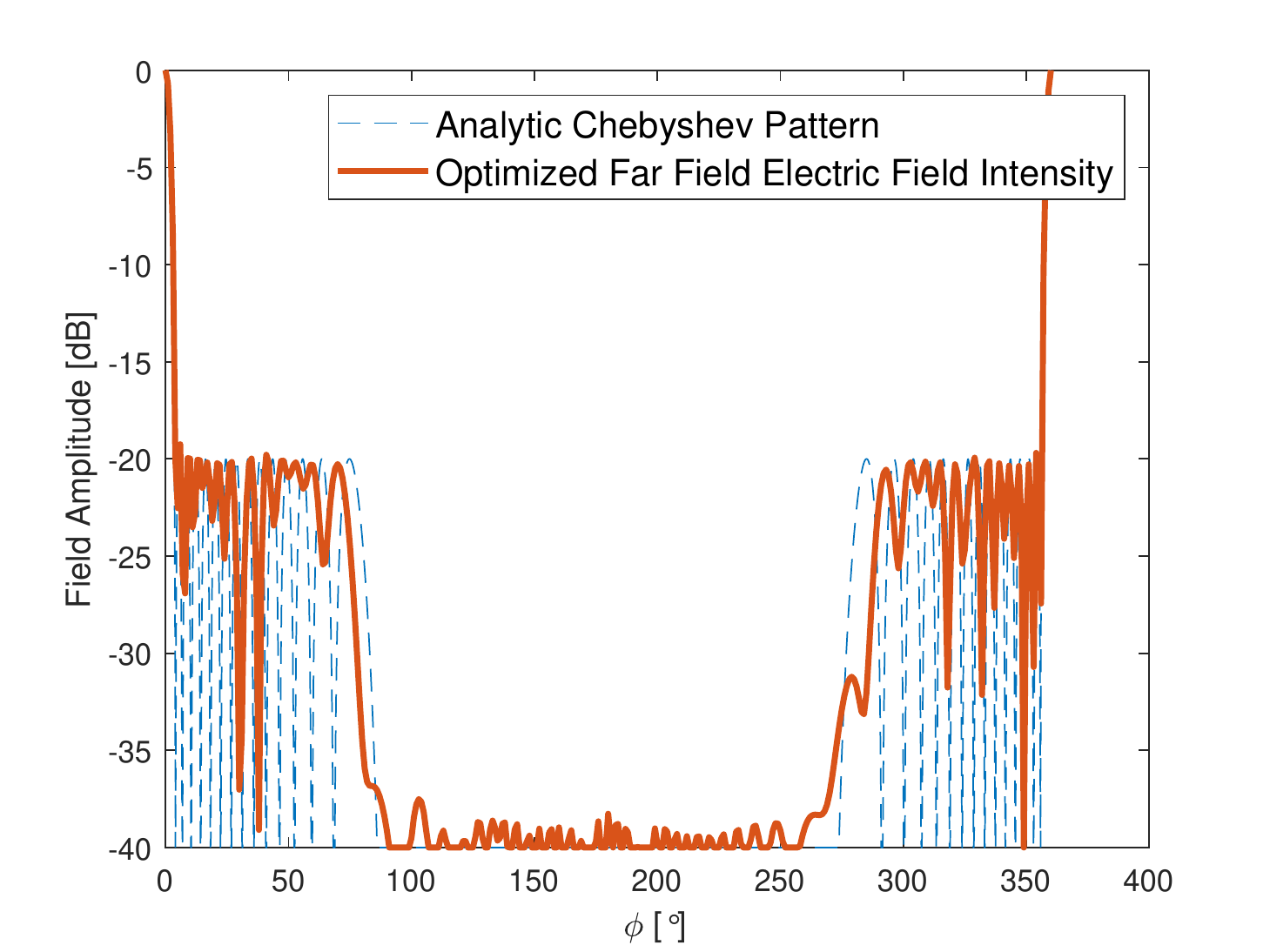}
% where an -eps-converted-to.pdf filename suffix will be assumed under latex, 
% and a .pdf suffix will be assumed for pdflatex; or what has been declared
% via \DeclareGraphicsExtensions.
\caption{Comparison of Chebyshev pattern with directive elements and optimized EMMS for a side lobe level of -20dB}
\label{Cheby20}
\end{figure}

\begin{figure}[!t]
\centering
\includegraphics[width=3in]{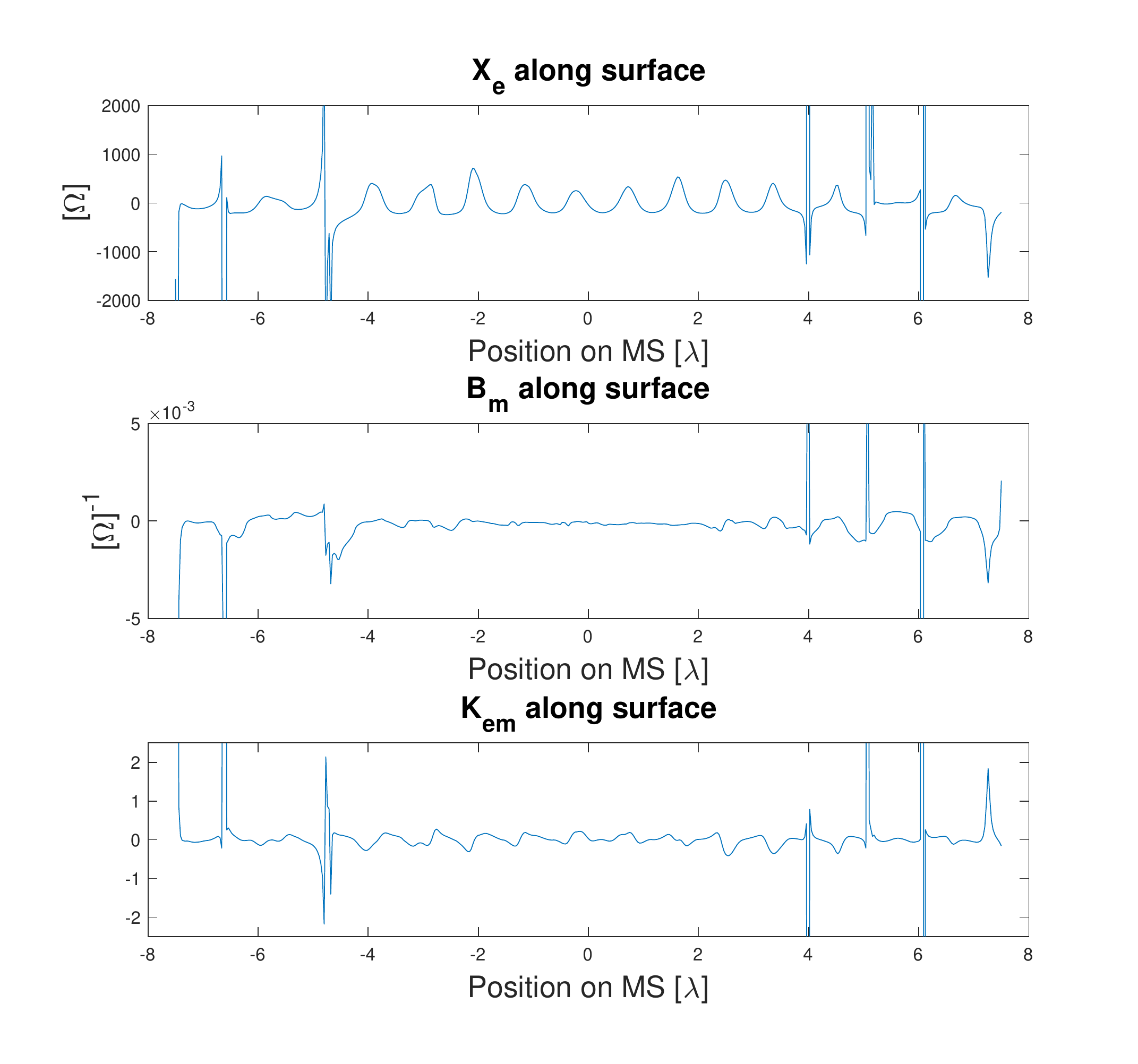}
% where an -eps-converted-to.pdf filename suffix will be assumed under latex, 
% and a .pdf suffix will be assumed for pdflatex; or what has been declared
% via \DeclareGraphicsExtensions.

% You want to know how I got these amazing insets? https://www.mathworks.com/matlabcentral/fileexchange/26007-on-figure-magnifier. Just run that function like this for subplots.
%h = subplot(1,3,3);
%magnifyOnFigure(h)
\caption{Surface parameters for the Chebyshev beamforming EMMS in \Cref{Cheby20}}
\label{Cheby20Loading}
\end{figure}

\begin{figure}[!t]
\centering
\includegraphics[width=2.75in]{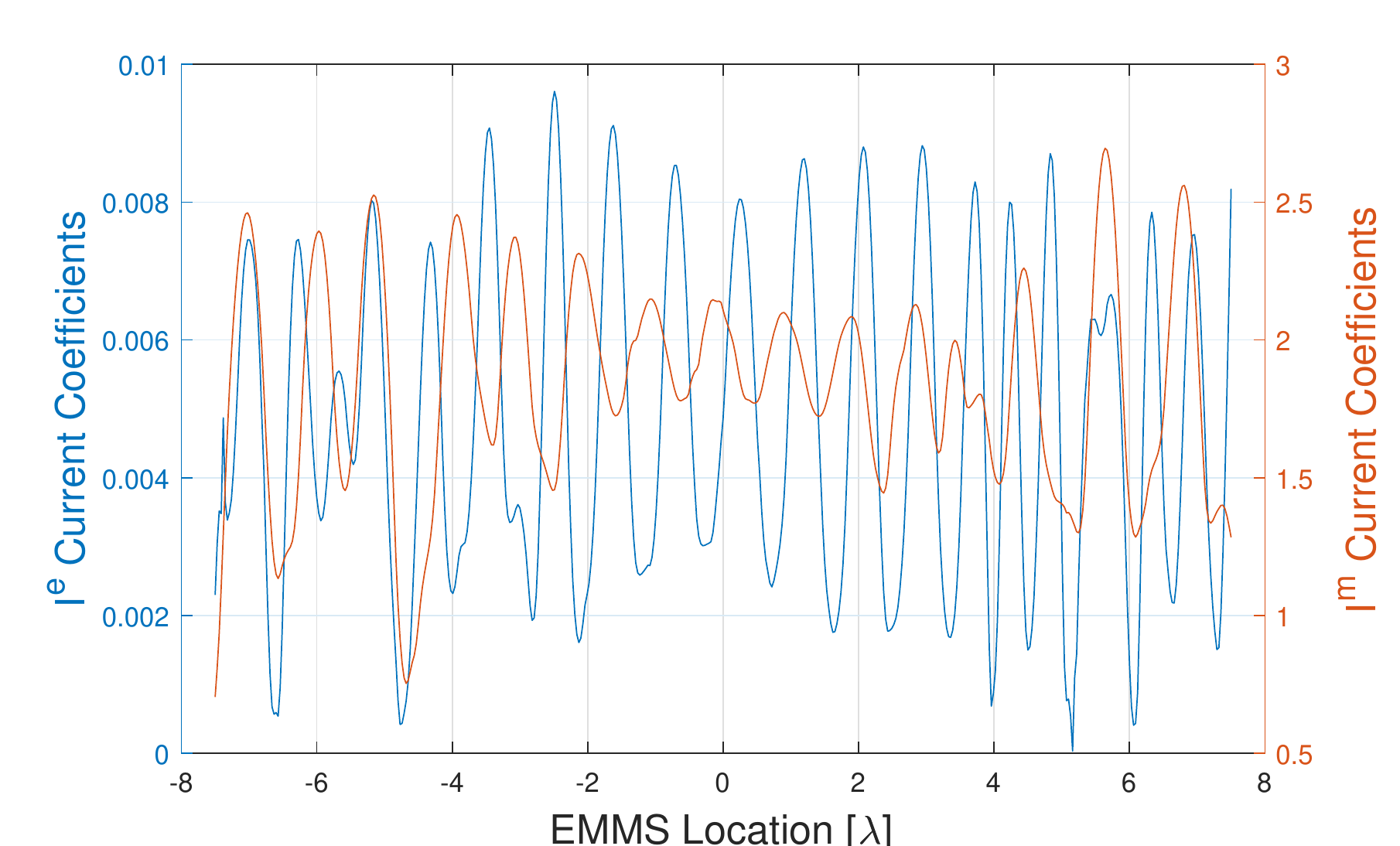}
% where an -eps-converted-to.pdf filename suffix will be assumed under latex, 
% and a .pdf suffix will be assumed for pdflatex; or what has been declared
% via \DeclareGraphicsExtensions.
\caption{Surface current coefficients for the Chebyshev beamforming EMMS in \Cref{Cheby20} }
\label{Cheby20Currents}
\end{figure}

The far-field criteria to satisfy is the Chebyshev side lobe level on the transmitted side with the appropriate beamwidth and -40dB on the reflected side ($\phi \in [90^{\circ},270^{\circ}]$). We remove constraints on the main beam level and null locations as they are not applicable. We have chosen a side lobe level of -20dB to realize as shown  in \Cref{Cheby20}. The far-field parameters to realize are listed in \Cref{Cheby20_table}. The currents and surface parameters for this example are shown in \Cref{Cheby20Currents} and \Cref{Cheby20Loading} respectively. As we can see, it satisfies the specified side lobe level far-field requirement, while also maintaining the HPBW of $3.7^{\circ}$. 

\begin{table}[!t]
\renewcommand{\arraystretch}{1.3}
\caption{-20dB Chebyshev Optimization Parameters}
\label{Cheby20_table}
\centering
\begin{tabular}{c|c|c|c}
\hline
\bfseries Side Lobe & \bfseries $D^e_{max}$  & \bfseries $D^m_{max}$ & \bfseries $\left\lbrace \alpha_{SL},\alpha_{D}\right\rbrace$ \\
\hline\hline
$\left\lbrace  4^{\circ}:356^{\circ} \right\rbrace  $ & 25 & 0.1 & $\left\lbrace 50,1 \right\rbrace$ \\
\hline
\end{tabular}
\end{table}


\section{Conclusion}
In this paper we introduced a new method for solving for EMMS surface parameters. The method uses a MoM model to fully incorporate edge effects and mutual coupling. Using this MoM model, a convex optimization-based solver using ADMM is implemented to solve for surface parameters based on far field criteria. These criteria are main beam location and level, null location, and maximum allowable side lobe level. Three examples were then presented to demonstrate this method. These are a multi-criteria example demonstrating all of the criteria at once, extreme angle refraction with electrically small EMMSs, and an example demonstrating Chebyshev beamforming. 

Although this method has many potential uses due to its flexible MoM model and varied far-field criteria, there remain some areas for future work. The main area for future research would be a method to determine the feasibility of far field criteria for a certain EMMS. We know heuristically that a certain sized EMMS can only produce so much directivity but these restrictions become less clear when we introduce more sophisticated far-field criteria. A way to determine the feasibility of far field criteria for an EMMS would be valuable. In addition, we have not been able to fashion a convex constraint for a minimum beam level. This would be a valuable capability, which would enable true mask-based constraints with a minimum and maximum level.

% or


%\appendices
%% you can choose not to have a title for an appendix
%% if you want by leaving the argument blank
%\section{}

%\section*{Acknowledgment}

% by thEMMSelves when using endfloat and the captionsoff option.
\ifCLASSOPTIONcaptionsoff
  \newpage
\fi



\bibliographystyle{IEEEtran}
\bibliography{IEEEabrv,./library}
\end{document}